\documentclass{elsart}

\usepackage{graphicx}

\usepackage{amssymb}

\begin{document}

\begin{frontmatter}
\title{Measurement-induced disturbance and thermal entanglement in spin models}
\author[label1,label2]{Guo-Feng Zhang\corauthref{cor1}},
\corauth[cor1]{Corresponding author:gf1978zhang@buaa.edu.cn}
\author[label3]{Zhao-Tan Jiang}
\author[label4]{and Ahmad Abliz}
\address[label1]{Department of Physics, School of Physics and Nuclear Energy Engineering, Beijing University of Aeronautics and Astronautics, Xueyuan Road No. 37, Beijing 100191, P. R. China}
\address[label2]{Department of Physics, Oklahoma State University, Stillwater, OK 74078, USA}
\address[label3]{Department of Physics, Beijing Institute of Technology, Beijing 100081, P. R. China}
\address[label4]{School of Mathematics, Physics and Informatics, Xinjiang Normal University, Urumchi 830054,
P. R. China}
\begin{abstract}
\quad Quantum correlation in two-qubit spin models is investigated by use of measurement-induced disturbance [Phys. Rev. A, 77, 022301 (2008)]. Its dependences on external magnetic field, spin-spin coupling, and Dzyaloshinski-Moriya (DM) interactions are presented in detail. We also compare measurement-induced disturbance and thermal entanglement in spin models and illustrate their different characteristics.
\end{abstract}

\end{frontmatter}

\section{Introduction}
It is well known that there are two fascinating features in quantum physics: The first is quantum entanglement which plays a central role in quantum information processing and there has been an ongoing effort to characterize entanglement qualitatively and quantitatively in recent years, and the second is quantum correlation arising from noncommutativity of operators representing states, observables, and measurements \cite{sll}. Quantum entanglement can be realized in many kinds of physical systems which involve quantum correlation.

Many studies concentrate on entanglement properties of condensed matter systems and application in quantum communication and information. An important emerging field is the quantum entanglement in solid state systems such as spin chains. Spin chains are the natural candidates for the realization of the entanglement compared with the other physical systems. The spin chains not only have useful applications such as the quantum state transfer, but also display rich entanglement features \cite{sbo}. The Heisenberg chain, the simplest spin chain, has been used to construct a quantum computer and quantum dots \cite{dlo}. By suitable coding, the Heisenberg interaction alone can be used for quantum computation \cite{dal,dpd,lfs}. The thermal entanglement, which requires neither measurement nor controlled switching of interactions in the preparing process, becomes an important quantity of systems for the purpose of quantum information. A lot of interesting work about thermal entanglement have been done \cite{man,xwa,glk,kmo,gfz1,yye,dvk,gfz2,xgw,shi,mca}.

But quantum correlation seems to have been seldom exploited before, especially for the solid spin system. Even many people take it for granted that quantum entanglement is quantum correlation. The entanglement properties of pure state are solved, but how to measure and classify entanglement for a mixed state remain unclear although some ingenious methods have been developed. In particular, some elegant analytical formulas for the entanglement of formation have been found for any two-qubit states \cite{shi}. Nevertheless, Li and Luo illustrated through simple examples that the entanglement of formation may exceed the total correlations as quantified by the quantum mutual information \cite{nli}. Several authors have pointed out
that in some quantum tasks which cannot be simulated by classical methods, it is the correlations (of course, of a quantum nature), rather than entanglement, that are responsible for the improvements.

An alternative classification for correlations based on quantum measurements has arisen in recent years \cite{mpi,slu,nli2}. In particular, quantum discord (QD) as a measure of quantum
correlations, initially introduced by Ollivier and Zurek \cite{hol}
and by Henderson and Vedral \cite{lhe} is attracting increasing
interest \cite{jcu,mdl,jsx,gad,rva,pgi,rcg,dos,mzw}. The QD for a large family of two-qubit states has been evaluated analytically, the authors have made a comparative study of the relationships between classical and quantum correlations in terms of QD, also and the entanglement of formation \cite{ssl2}. For the general quantum correlations and their classical
counterparts, under the action of decoherence, Maziero \emph{et al.} \cite{jma} identify three general types of dynamics that include a peculiar sudden change in their decay rates. They show, under suitable conditions, the classical correlation
is unaffected by decoherence. Most of the above works are limited to studies of bipartite correlations only as the concept of
discord, which relies on the definition of mutual information,
is not defined for multipartite systems. In some of the
studies, it is also desirable to compare various notions of
quantum correlations. Kavan \emph{et al.} \cite{kmod} discuss the problem of the separation of total correlations in a given quantum state into entanglement, dissonance, and classical correlations using the concept of relative entropy as a distance
measure of correlations. Their results show that dissonance may be present in pure multipartite states. Mazzola \emph{et al.} \cite{lma} find a sudden transition from classical to quantum decoherence regime by studying the dynamics of classical and quantum correlations (quantified by QD) in the presence of nondissipative decoherence. Recently, some authors \cite{twe} have pointed out that thermal quantum discord (TQD), in contrast to entanglement and other thermodynamic quantities, spotlight the critical points associated with quantum phase transitions (QPTs) for some spin chain model even at finite $T$. They think that the remarkable property of TQD is an important
tool that can be readily applied to reduce the experimental
demands to determine critical points for QPTs. More recently, Luo \emph{et al}. evaluate the geometric measure of QD \cite{dak} for an arbitrary state and obtain an explicit and tight lower bound, their results show QD actually coincides with a simpler quantity based on von Neumann measurements \cite{ssl3}. They think their simple and explicit bound makes QD a convenient and interesting tool for analyzing quantum correlations.

Now we recognize that quantum entanglement is a special kind of quantum correlation, but not the same with quantum correlation. So, it is very interesting and necessary to study the relation between quantum entanglement and quantum correlation. Moreover, the effects of external parameters on quantum correlation in spin chain also deserve to be investigated. Luo \cite{sll} introduced a classical vs quantum dichotomy in order to classify and quantify statistical correlations in bipartite states. In this paper, we will explore quantum correlation based on Luo's method \cite{sll} and investigate the dependences of spin-spin coupling, DM interaction, temperature and external magnetic field on quantum correlation in two two-qubit models. The comparison between quantum correlation and thermal entanglement measured by concurrence will be given.

The paper is organized as follows. In Sec. 2, we recall thermal state, entanglement and quantum correlation measured by concurrence and measurement-induced disturbance \cite{sll}, respectively. In Sec. 3, we will investigate measurement-induced disturbance and quantum entanglement in two two-qubit spin models and give a detailed comparison. The effects of spin-spin coupling, DM interaction, temperature and external magnetic field on the two prominent characteristics of quantum physics are illustrated. Finally, Sec. 4 is devoted to conclusions.

\section{Thermal state, Quantum correlation via measurement-induced disturbance and quantum thermal entanglement measured by concurrence}
\emph{Thermal state.} For a system which can be described by $H$ in equilibrium at temperature $T$, the density matrix is $\rho(T)=e^{-\beta\emph{H}}/Z$, where $\beta=1/(k\emph{T})$ , $k$ is the Boltzmann constant and
$Z=$tr$e^{-\beta\emph{H}}$ is the partition function. For
simplicity, we write $k=1$. Due to the introduction of temperature, we call $\rho$ thermal state.

\emph{Quantum correlation via measurement-induced disturbance.} We can apply local measurement $\{\prod_{k}\}$($\prod_{k}\prod_{k^{'}}=\delta_{kk^{'}}\prod_{k}$ and $\sum_{k}\prod_{k}=1$) to any bipartite state $\rho$ (of course, including thermal state). Here $\prod_{k}=\prod_{i}^{a}\otimes\prod_{j}^{b}$ and $\prod_{i}^{a}$, $\prod_{j}^{b}$ are complete projective measurements consisting of one-dimensional orthogonal projections for parties $a$ and $b$. After the measurement, we get the state $\prod(\rho)=\sum_{ij}(\prod_{i}^{a}\otimes\prod_{j}^{b})\rho(\prod_{i}^{a}\otimes\prod_{j}^{b})$ which is a classical state \cite{sll}. If the measurement
$\prod$ is induced by the spectral resolutions of the reduced states $\rho^{a}=\sum_{i}p_{i}^{a}\prod_{i}^{a}$ and $\rho^{b}=\sum_{j}p_{j}^{b}\prod_{j}^{b}$, the measurement leaves the marginal information invariant and is in a certain sense the least disturbing. In fact, $\prod(\rho)$ is a classical state that is closest to the original state $\rho$ since this kind of measurement can leave the reduced states invariant. One can use any reasonable distance between $\rho$ and $\prod(\rho)$ to measure the quantum correlation in $\rho$. In this paper, we will adopt Luo's method \cite{sll}, i.e., quantum mutual information difference between $\rho$ and $\prod(\rho)$,  to measure quantum correlation in $\rho$. The total correlation in a bipartite state $\rho$ can be well quantified by the quantum mutual information $I(\rho)=S(\rho^{a})+S(\rho^{b})-S(\rho)$, and $I(\prod(\rho))$ quantifies the classical correlations in $\rho$ since $\prod(\rho)$ is a classical state. Here $S(\rho)=$-tr$\rho$log$\rho$ denotes the von Neumann entropy, and the logarithm is always understood as base $2$ in this paper. So the quantum correlation can be quantified by the measurement-induced disturbance\cite{sll}
\begin{equation}
Q(\rho)=I(\rho)-I(\prod(\rho)).
\end{equation}

\emph{Quantum thermal entanglement measured by concurrence.} The entanglement of two qubits state $\rho$ (of course, including thermal state) can be measured by the concurrence $C(\rho)$ which is written as $C(\rho)=\max[0,2 \max[\lambda_{i}]-\sum^{4}_{i}\lambda_{i}]$\cite{shi}, where
$\lambda_{i}$ are the square roots of the eigenvalues of the
matrix $R=\rho S\rho^{*}S$, $\rho$ is the density matrix,
$S=\sigma_{1}^{y}\otimes\sigma_{2}^{y}$ and $*$ stands for the
complex conjugate. The concurrence is available no matter whether $\rho$ is pure or mixed. We term the entanglement, which measured by concurrence, associated with thermal state $\rho(T)$ as thermal concurrence.

\section{Quantum correlation characterized by measurement-induced disturbance in two two-qubit spin models}

In this section, we will investigate quantum correlation in a two-qubit Heigenberg \emph{XXZ} spin model under an inhomogeneous magnetic field and in a two-qubit \emph{XXX} spin model with DM anisotropic antisymmetric interaction. Moreover, we also study thermal concurrence and give a comparison with quantum correlation. The effects of spin-spin coupling, DM interaction, temperature and external magnetic field on the two prominent characteristics of quantum physics are illustrated.

\subsection{Quantum correlation in an XXZ spin model}

We consider the model whose Hamiltonian can be described with
\begin{equation}
\emph{H}=\frac{1}{2}[J(\sigma_{1}^{x}\sigma_{2}^{x}+\sigma_{1}^{y}\sigma_{2}^{y})
+J_{z}\sigma_{1}^{z}\sigma_{2}^{z}\nonumber\\ +(B+b)\sigma_{1}^{z}+(B-b)\sigma_{2}^{z}],
\end{equation}
where $J$ and $J_{z}$ are the real coupling coefficients and $\sigma_{i}^{\alpha}(\alpha=x,y,z; i=1,2)$ are Pauli spin operators. $B\geq0$ is restricted, and the magnetic fields on the two spins have been so parameterized that \emph{b} controls the degree of inhomogeneity. Note that we are working in units so that $B$, $b$, $J$ and $J_{z}$ are dimensionless. The thermal concurrence has been studied in Ref.\cite{gfz1}, here we mainly focus on quantum correlation and the comparison between quantum correlation and thermal concurrence.

We can obtain thermal state $\rho(T)=e^{-\beta\emph{H}}/Z$, in the standard basis
$\{|1,1\rangle,|1,0\rangle,\\|0,1\rangle,|0,0\rangle\}$, which can be expressed as
\begin{eqnarray}
\rho=\frac{1}{Z}
\left(%
\begin{array}{cccc}
  e^{-\frac{J_{z}+2B}{2T}}& 0  &0 & 0 \\
  0 & \rho_{22} & -s & 0 \\
  0 &  -s & \rho_{33} & 0\\
 0 &0 & 0 & e^{-\frac{J_{z}-2B}{2T}} \\
\end{array}%
\right),
\end{eqnarray}
where
$\rho_{22}=e^{J_{z}/(2T)}(m-n)$, $\rho_{33}=e^{J_{z}/(2T)}(m+n)$, $m=\cosh[\eta/T]$, $n=b\sinh[\eta/T]/\eta$, $\eta=\sqrt{b^{2}+J^{2}}$, $s=e^{J_{z}/(2T)}J\sinh[\eta/T]/\eta$, and $Z=(1+e^{2B/T}+2me^{(J_{z}+B)/T})e^{-(J_{z}+2B)/(2T)}$. We have written the Boltzmann constant $k=1$. Note that the spectra of $\rho$ consist of $e^{-\frac{J_{z}\pm2B}{2T}}/Z$,  $e^{\frac{J_{z}\pm2\eta}{2T}}/Z$, and
\begin{eqnarray}
\rho^{a}=\frac{1}{Z}
\left(%
\begin{array}{cccc}
  e^{-\frac{J_{z}+2B}{2T}}+\rho_{22}&0 \\
   0&\rho_{33}+e^{-\frac{J_{z}-2B}{2T}}\\
\end{array}%
\right),
\end{eqnarray}
\begin{eqnarray}
\rho^{b}=\frac{1}{Z}
\left(%
\begin{array}{cccc}
  e^{-\frac{J_{z}+2B}{2T}}+\rho_{33}&0 \\
   0&\rho_{22}+e^{-\frac{J_{z}-2B}{2T}}\\
\end{array}%
\right).
\end{eqnarray}
On the other hand, by taking $\prod_{i}^{a}=|i\rangle\langle i|$ and $\prod_{j}^{b}=|j\rangle\langle j|$, we have
\begin{eqnarray}
\prod(\rho)=\frac{1}{Z}
\left(%
\begin{array}{cccc}
  e^{-\frac{J_{z}+2B}{2T}}& 0  &0 & 0 \\
  0 & \rho_{22} & 0 & 0 \\
  0 &  0 & \rho_{33} & 0\\
 0 &0 & 0 & e^{-\frac{J_{z}-2B}{2T}} \\
\end{array}%
\right),
\end{eqnarray}
and
\begin{eqnarray}
[\prod(\rho)]^{a}=\rho^{a}, [\prod(\rho)]^{b}=\rho^{b}.
\end{eqnarray}
Consequently,
\begin{eqnarray}
Q(\rho)&=&-\frac{\rho_{22}}{Z}\log\frac{\rho_{22}}{Z}-\frac{\rho_{33}}{Z}\log\frac{\rho_{33}}{Z}\nonumber\\&+&\frac{e^{\frac{J_{z}-2\eta}{2T}}}{Z}\log\frac{e^{\frac{J_{z}-2\eta}{2T}}}{Z}+\frac{e^{\frac{J_{z}+2\eta}{2T}}}{Z}\log\frac{e^{\frac{J_{z}+2\eta}{2T}}}{Z}.
\end{eqnarray}

Also, we can obtain thermal concurrence associated with Eq.(3)
\begin{eqnarray}
C(\rho)=\max\{\frac{\sqrt{\xi_{+}}}{\eta Z^{2}}-\frac{\sqrt{\xi_{-}}}{\eta Z^{2}}-\frac{2e^{\frac{-J_{z}}{2T}}}{Z},0\},
\end{eqnarray}
with $\xi_{\pm}=e^{J_{z}/T}Z^{2}\mu\pm\sqrt{2}\nu$, $\nu=\sqrt{J^{2}Z^{4}(b^{2}+\mu+J^{2})\sinh[\eta/T]^{2}}e^{J_{z}/T}$, and $\mu=b^{2}+J^{2}\cosh[2\eta/T]$.
\begin{figure}[h]
\begin{center}
\includegraphics[width=10 cm]{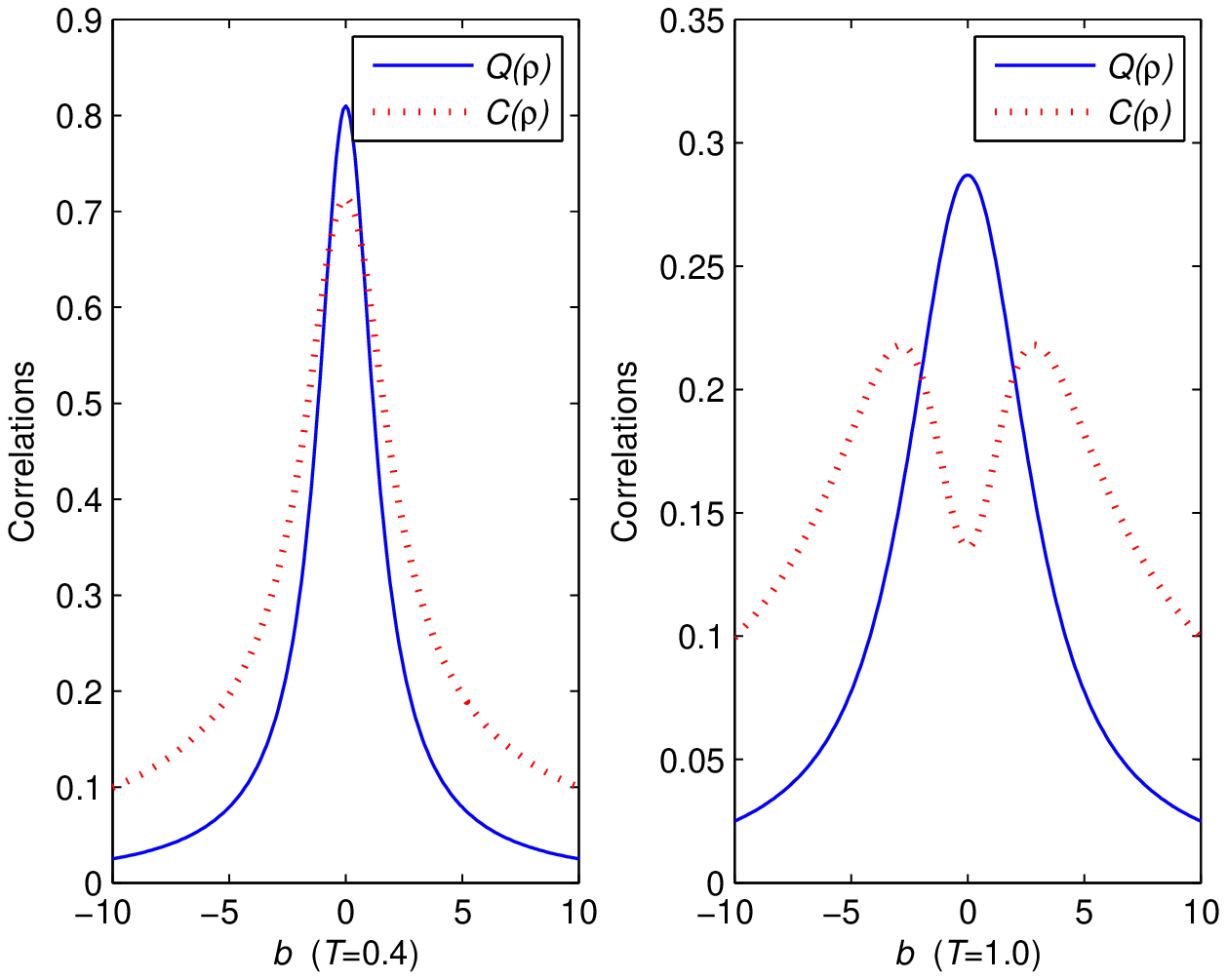}
\caption{(Color online) Quantum correlation and thermal concurrence for $B=0$,  $J_{z}=0$ and $J=1$ case. $T$ is plotted in units of the Boltzmann constant $k$. And we work in units where $B$ and $b$ are dimensionless.}
\includegraphics[width=11 cm,height=6 cm]{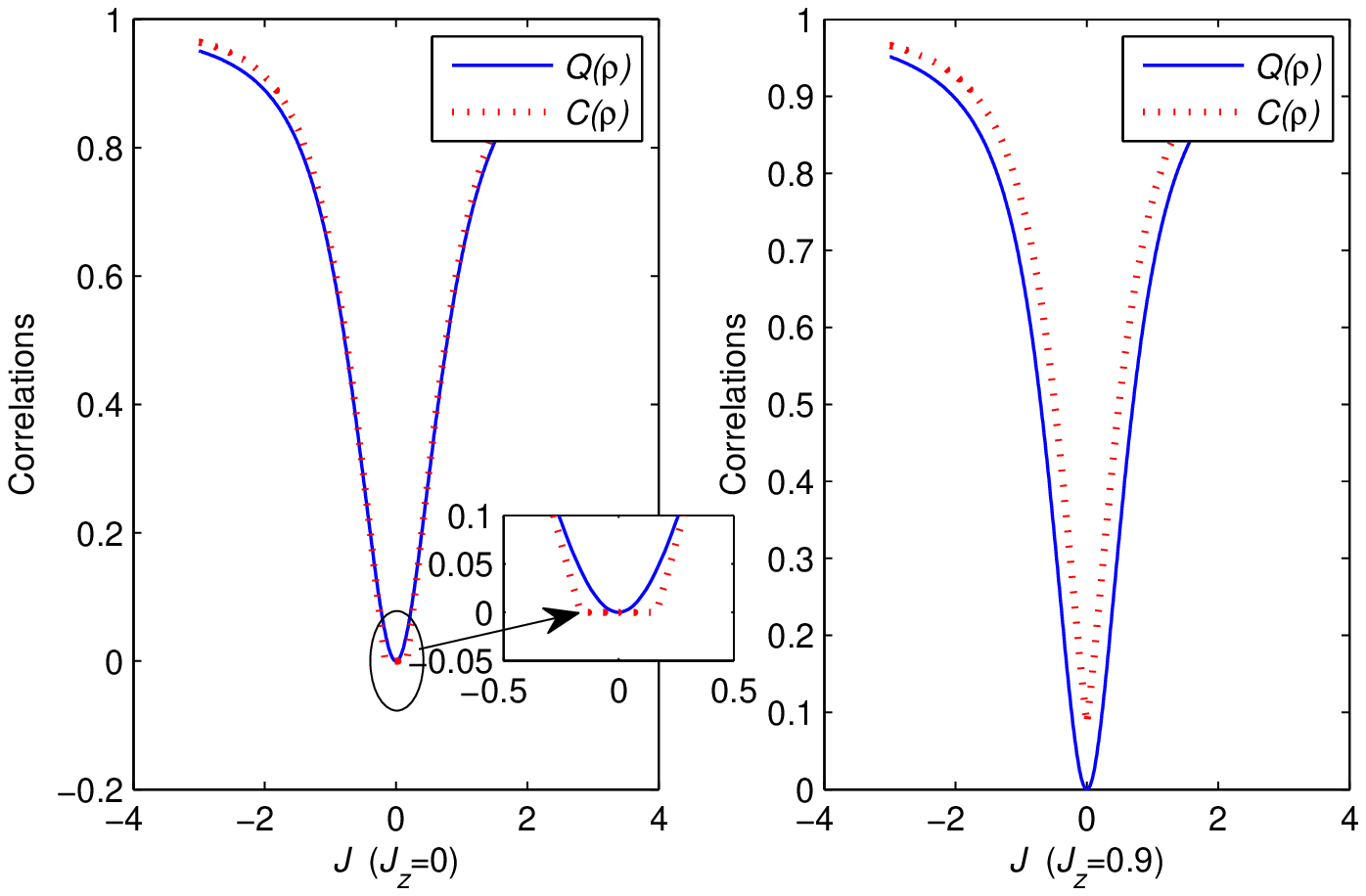}
\caption{(Color online) Quantum correlation and thermal concurrence for $B=0$,  $b=0.8$ and $T=0.4$ case. $T$ is plotted in units of the Boltzmann constant $k$. And we work in units where $B$ and $b$ are dimensionless.}
\end{center}
\end{figure}

\emph{Case1: $J_{z}=0$.} Our model corresponds to an \emph{XX} spin
model. We know that in any solid state construction of qubits, there is always the possibility of inhomogeneous Zeeman coupling. So it is necessary to investigate quantum correlation and thermal concurrence's dependences on nonuniform magnetic field.
In Fig. 1, we give the results at different temperatures for the nonuniform magnetic field ($B=0$).
From the figures, we can know that the two quantities are symmetric with respect to the zero magnetic field,
the nonuniform magnetic field can lead to lower quantum correlation. Quantum correlation evolves alike and decreases monotonously with $|b|$ at any temperature, while thermal concurrence has a double-peak structure at a higher temperature.
\begin{figure}
\begin{center}
\includegraphics[width=11 cm,height=6 cm]{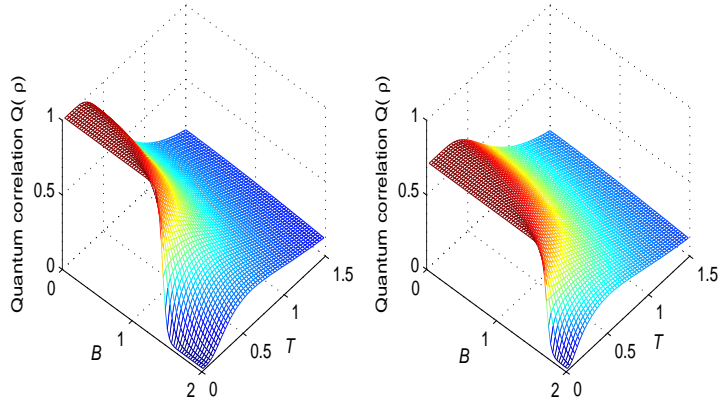}
\caption{(Color online) Quantum correlation is plotted vs \emph{B} and \emph{T}. Coupling constant
$J=1$, $J_{z}=0.4$ and magnetic field $b=0$ (left), $b=0.8$ (right). $T$ is plotted in units of the Boltzmann constant $k$. And we work in units where $B$ is dimensionless.}
\end{center}
\end{figure}
\begin{figure}
\begin{center}
\includegraphics[width=11 cm,height=6 cm]{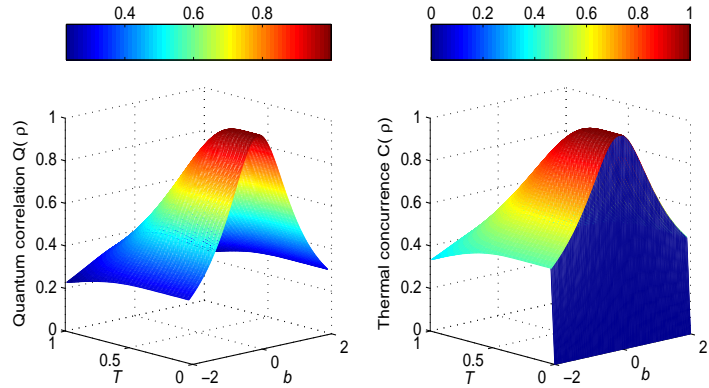}
\caption{(Color online) Quantum correlation and thermal concurrence are plotted vs \emph{b} and \emph{T}. Coupling constant $J=1$, $J_{z}=0.8$ and magnetic field $B=0$. $T$ is plotted in units of the Boltzmann constant $k$. And we work in units where $b$ is dimensionless.}
\end{center}
\end{figure}

\emph{Case2: For any $J_{z}$.} In Fig.2, the plots of these two quantities with respect to spin-spin coupling $J$ for different $J_{z}$ under a nonuniform magnetic field are given. It is can be seen that the introducing of $J_{z}$ can excite more
quantum correlation and thermal concurrence. For ferromagnetic $(J<0)$ and anti-ferromagnetic $(J>0)$,
these two quantities exist and are symmetric about $J$.
However, if $b=0$, and $J_{z}=J$, i.e., the model changes to $XXX$ type,
quantum correlation and thermal concurrence will be not symmetric about $J$, this
can be seen from Fig.6 in Sec. 3. 2. Thermal concurrence will
be zero when $|J|$ is smaller, while quantum correlation is
always non-vanishing. A rather counterintuitive feature of the thermal concurrence is that it may exceed the quantum correlation (see the right panel of Fig.2). So one can say quantum correlation is
more general than thermal concurrence. In order to see the
effects of temperature and uniform magnetic field $B$,
quantum correlation is plotted versus \emph{B} and \emph{T}
in Fig.3. Temperature will make quantum correlation be weaker,
which can be easily understood since it belongs to
quantum character. Uniform magnetic field $B$ also plays a
negative role for a fixed temperature. Temperature can play a positive role
when external magnetic field is strong, for example,
$B=2$ in Fig.3, quantum correlation increases firstly with
temperature although it will fall to zero eventually at larger
temperature. However, when $b$ is raised, the critical
magnetic field $B_{c}$ (above which quantum correlation is vanishing)
increases, but the maximum value at which the system
can arrive at becomes smaller which is the same with thermal
concurrence \cite{gfz1}. It is shown that thermal concurrence
experiences a sudden transition when temperature changes
from a finite value to zero for any nonuniform magnetic field in the right panel of Fig. 4,
while quantum correlation evolves continuously with respective
to temperature even it tends to be zero.

\subsection{Quantum correlation in an XXX spin model with DM anisotropic antisymmetric interaction}

Next, we consider the following model
\begin{equation}
H_{DM}=\frac{J}{2}[(\sigma_{1}^{x}\sigma_{2}^{x}+\sigma_{1}^{y}\sigma_{2}^{y}+\sigma_{1}^{z}\sigma_{2}^{z})+\overrightarrow{D}\cdot(\overrightarrow{\sigma_{1}}\times\overrightarrow{\sigma_{2}})],
\end{equation}
here $J$ is the real coupling coefficient and $\overrightarrow{D}$
is the DM vector coupling. The DM anisotropic antisymmetric
interaction arises from spin-orbit coupling \cite{idz,tmo}. For simplicity, we choose $\overrightarrow{D}=D\overrightarrow{z}$. In the standard basis
$\{|1,1\rangle,|1,0\rangle,|0,1\rangle,|0,0\rangle\}$,
thermal state $\rho(T)=e^{-\beta H_{DM}}/Z$ is
\begin{eqnarray}
\rho=\frac{1}{Z}
\left(%
\begin{array}{cccc}
  e^{-\frac{J}{2T}}& 0  &0 & 0 \\
  0 & \rho_{22} & -\frac{M_{-}L_{-}e^{i\theta}}{2} & 0 \\
  0 &  -\frac{M_{-}L_{-}e^{-i\theta}}{2} & \rho_{33} & 0\\
 0 &0 & 0 & e^{-\frac{J}{2T}} \\
\end{array}%
\right),
\end{eqnarray}
where $\delta=2J\sqrt{1+D^2}$, $Z=2e^{-J/(2T)}(1+e^{J/T}\cosh[\delta/(2T)])$, $\rho_{22}=\rho_{33}=L_{-}M_{+}/2$, and $L_{\pm}=e^{(J\pm\delta)/(2T)}$, $M_{\pm}=\pm1+e^{\delta/T}$. Using the same method as \emph{3.1}, quantum correlation and thermal concurrence can be obtained
\begin{equation}
Q(\rho)=-\frac{\rho_{22}}{Z}\log\frac{\rho_{22}}{Z}-\frac{\rho_{33}}{Z}\log\frac{\rho_{33}}{Z}+\frac{L_{-}}{Z}\log\frac{L_{-}}{Z}+\frac{L_{+}}{Z}\log\frac{L_{+}}{Z},
\end{equation}
\begin{eqnarray}
C(\rho)=\max\{\frac{2}{Z}(\frac{1}{2}|L_{-}M_{-}|-e^{-\frac{J}{2T}}),0\}.
\end{eqnarray}

\emph{Case1: $D=0$.} Eq.(12) and Eq.(13) can be reduced to
\begin{eqnarray}
Q(\rho)=\frac{4x^{4}\log x-(1+x^{4})(-1+\log(1+x^{4}))}{3+x^{4}},
\end{eqnarray}
\begin{eqnarray}
C(\rho)=\max\{\frac{-2+|-1+x^{4}|}{3+x^{4}},0\},
\end{eqnarray}
with $x=e^{J/(2T)}$. Quantum correlation exists for both antiferromagnetic and ferromagnetic case. Both thermal concurrence and quantum correlation will tend to be $1$ when $x$ approaches positive infinity. Moreover, quantum correlation increases monotonically from zero with $x$ in $x>1$ region and will be $1/3$ when $x=0$, i.e., $J$ is negative infinity or the temperature is very
high while there is no thermal concurrence when $x<3^{1/4}\simeq1.316$.
These can be seen from Fig.5.
\begin{figure}
\begin{center}
\includegraphics[width=9 cm,height=6 cm]{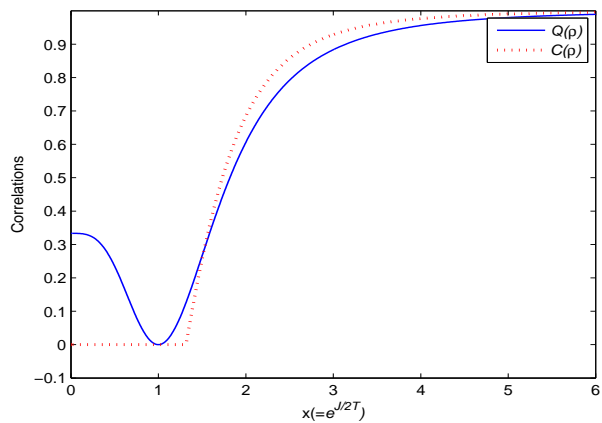}
\caption{Quantum correlation and thermal concurrence are plotted vs \emph{x}($=e^{\frac{J}{2T}}$) for $D=0$.}
\end{center}
\end{figure}

\emph{Case2: For any $D$.} Fig.6 show quantum correlation and thermal concurrence are not symmetric with respective to spin-spin coupling $J$ and
antiferromagnetic coupling can be more helpful for entanglement. There is no thermal concurrence for a ferromagnetic \emph{XXX} model when DM interaction is weak while quantum correlation exists for any DM interaction.
\begin{figure}
\begin{center}
\includegraphics[width=11 cm,height=6 cm]{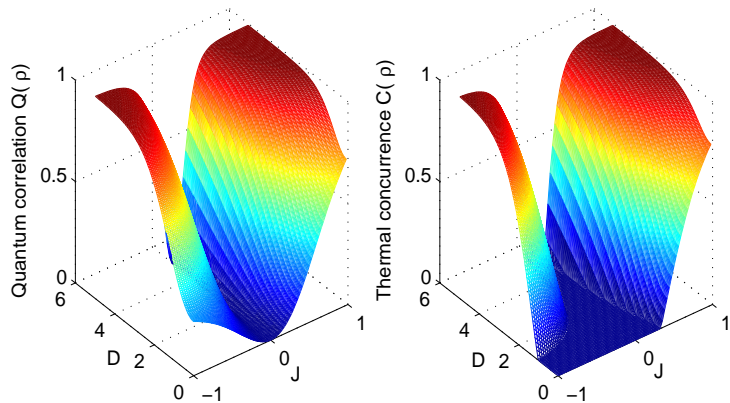}
\caption{(Color online) Thermal concurrence and quantum correlation are plotted vs \emph{D} and \emph{J} for $T=0.6$.}
\end{center}
\end{figure}

\section{Conclusions } In the past studies, quantum entanglement is investigated widely. However, quantum entanglement is only a special kind of quantum correlation. Many people take it for granted that quantum correlation exists only in entangled state. By using measurement-induced disturbance we have investigated quantum correlation in two two-qubit spin model. The dependences of measurement-induced disturbance on external magnetic field, spin-spin coupling and DM anisotropic antisymmetric interaction are given in detail. More important, we have compared measurement-induced disturbance with quantum thermal entanglement measured by concurrence and found no definite link between them. However, the effect of temperature on measurement-induced disturbance is far weaker than on thermal concurrence. At low temperature, for an \emph{XX} model without  magnetic field ($B=0$ and $b=0$), thermal concurrence and measurement-induced disturbance arrive their peak values, but when the temperature is higher thermal concurrence will disappear while measurement-induced disturbance experiences a maximum value. Only a higher spin-spin coupling $J$ can excite thermal concurrence for an \emph{XX} model, but measurement-induced disturbance will exist for any $J$. Both of them will be increased by introducing $z$-direction coupling, especially for thermal concurrence. Also, $z$-direction coupling will raise critical uniform magnetic field of these two quantities. Interestingly, thermal concurrence will experience a sudden transition when temperature approaches zero, but this will not happen for measurement-induced disturbance. There is no thermal concurrence for a ferromagnetic \emph{XXX} model when DM interaction is weak, measurement-induced disturbance exists for both antiferromagnetic and ferromagnetic case. Both the thermal concurrence and measurement-induced disturbance will tend to be $1$ when $x(=e^{J/(2T)})$ approaches positive infinity. Moreover, measurement-induced disturbance increases monotonically from zero with $x$ when $x>1$ and will be $1/3$ when $x=0$, i.e. $J$ is negative infinity or temperature is very high while there is no thermal concurrence when $x<3^{1/4}\simeq1.316$. All results show that there is no direct relation between measurement-induced disturbance and quantum entanglement, separable state can possess quantum correlation.

\section{Acknowledgements}
This work was supported by the National Science Foundation of China under Grants No. 10874013, 10604053 and US-DOE under Grants No. DEFG02-04ER46124. Guo-Feng Zhang would like to thank Professor Xin-Cheng Xie (Department of Physics, Oklahoma State University,
USA) for the hospitality and the financial support. Zhao-Tan Jiang also acknowledges the support of the National Science Foundation of China under Grant No. 10974015. Ahmad Abliz acknowledges the support of the National Science Foundation of China under Grant No. 10664004.

\end{document}